
\tolerance=10000
\input phyzzx

\font\mybb=msbm10 at 12pt
\def\bb#1{\hbox{\mybb#1}}
\def\Z {\bb{Z}}
\def\R {\bb{R}}

\REF\Dbrane{R.G. Leigh, Mod. Phys. Lett. {\bf A28 }  (1989) 2767;
M. Douglas, hep-th/9512077; C. Schmidhuber, hep-th/9601003; M.B. Green, C.M. Hull
and P.K.
Townsend, Phys. Lett. {\bf B382} (1996) 65.}
\REF\polch{S. Chaudhuri, C. Johnson, and J. Polchinski,
``Notes on D-branes,'' hep-th/9602052; J. Polchinski,
``TASI Lectures on D-branes,'' hep-th/9611050}
\REF\Mat{T. Banks, W. Fischler, S. Shenker, and L. Susskind,
``M theory as a matrix model: a conjecture,''
hep-th/9610043, Phys. Rev. 
{\bf D55} (1997) 5112.}%
\REF\brev{For a review see T. Banks,
hep-th/9706168.}%
\REF\taylor{W. Taylor, hep-th/9611042, Phys. Lett. {\bf B394} (1997) 283.}%
\REF\motl{L. Motl,hep-th/9701025.}
\REF\IIAbs{T. Banks and N. Seiberg,hep-th/9702187.}
\REF\IIAVV{R. Dijkgraaf,E. Verlinde and H. Verlinde, hep-th/9703030.}
\REF\fhrs{W. Fischler, E. Halyo, A. Rajaraman and L. Susskind, 
hep-th/9703102.}%
\REF\dvv{R. Dijkgraaf, E. Verlinde, and H. Verlinde,  hep-th/9603126;
 hep-th/9604055; hep-th/9703030; hep-th/9704018.} 
\REF\rozali{M. Rozali,hep-th/9702136.}
\REF\brs{M. Berkooz, M. Rozali, and N. Seiberg, 
hep-th/9704089}
\REF\MatTor{N. Seiberg, 
hep-th/9705221.}%
\REF\mem{A. Losev, G. Moore and S.L. Shatashvili, hep-th/9707250; I. Brunner and A. Karch,
hep-th/9707259; A. Hanany and G. Lifschytz, hep-th/9708037. }
\REF\prog{C.M. Hull, in preparation}
\REF\Dis {D. Berenstein, R. Corrado and J. Distler, hep-th/9704087.}
\REF\Der{A.H. Bilge, T. Dereli and S. Kocak, Lett. Math. Phys. {\bf 36} (1996) 301; J. Math. Phys.
{\bf 38} (1997) 4804.}
\REF\Corr{E. Corrigan, C. Devchand, D.B.
 Fairlie and J. Nuyts, Nucl. Phys. {\bf B214} (1983) 452.}
\REF\SpinSdSol {D.B.
 Fairlie and J. Nuyts, J. Phys. {\bf A17} (1984) 431;
S. Fubini  and H. Nicolai, Phys. Lett. {\bf B155} (1985) 369; J. Harvey and A. Strominger,
 Phys. Rev. Lett. {\bf 66} (1991) 549. }
\REF\Pop{T.A. Ivanova and A.D. Popov, Lett. Math. Phys. {\bf 24} (1992) 85; Theor. Math. Phys.
{\bf 94} (1993) 225; T.A. Ivanova, Phys. Lett. {\bf B315} (1993) 277.}
\REF\Spingt {M. Gunaydin and H. Nicolai, Phys. Lett. {\bf B351} (1995) 169.}
\REF\Tset{I. Chepelev and A.A. Tseytlin, hep-th/9709087.}
\REF\Bob{B.S. Acharya, M. O'Loughlin and B. Spence, Nucl.
Phys. {\bf B503} (1997) 657; B.S. Acharya, J.M. Figueroa-O'Farrill, M. O'Loughlin and B. Spence,
hep-th/9707118; L. Baulieu, I. Kanno and I. Singer, hep-th/9704167.}
\REF\Bobb{
B.S. Acharya and M. O'Loughlin, Phys. Rev. {\bf D55} (1997) 353.}
\REF\sez{H. Nishino and  E. Sezgin, Phys. Lett. {\bf 388B} (1996) 569, hep-th/9607185; H. Nishino,
hep-th/9706148,9708064; E. Sezgin hep-th/9703123; I. Bars, hep-th/9704054; I. Bars and C. Kounnas,
hep-th/9612119.}
\REF\SSDU{S.J. Gates,  H. Nishino and S. V. Ketov, Phys. Lett. {\bf B267} (1992) 99;
Phys. Lett. {\bf B307} (1993) 323; Nucl. Phys. {\bf B393} (1993)  149;
E. Bergshoeff and E. Sezgin, Phys. Lett. {\bf
B292} (1992) 87.}


\def \aa {\alpha}

\def \dd {\delta}
\def \ee {\epsilon}

\def \ll {\lambda}
\def \mm {\mu}

\def \th {\theta}

 \def \ggg {\Gamma}

\def \lll {\wedge}

\def \ti {\tilde}

\def \2 {{1 \over 2}}
\def \3 {{1 \over 3}}
\def \4 {{1 \over 4}}
\def \5 {{1 \over 5}}
\def \6 {{1 \over 6}}
\def \7 {{1 \over 7}}
\def \8 {{1 \over 8}}
\def \9 {{1 \over 9}}
\def \0 { \infty}

\def\++ {{(+)}}
\def \- {{(-)}}
\def\+-{{(\pm)}}
 
\def\ek {\eqn\abc$$}


 \def\unit{\hbox to 3.3pt{\hskip1.3pt \vrule height 7pt width .4pt \hskip.7pt
\vrule height 7.85pt width .4pt \kern-2.4pt
\hrulefill \kern-3pt
\raise 4pt\hbox{\char'40}}}

\def\nup#1({Nucl.\ Phys.\  {\bf B#1}\ (}

\def\tr  {{\rm tr}}



\Pubnum{ \vbox{  \hbox {QMW-97-32} \hbox{LPTENS 97/47}  \hbox{hep-th/9710165}} }
\pubtype{}
\date{October, 1997}

\titlepage

\title {\bf  Higher Dimensional Yang-Mills Theories and Topological Terms}

\author{C.M. Hull}
\address{Physics Department, Queen Mary and Westfield College,
\break Mile End Road, London E1 4NS, U.K.}
\andaddress{Laboratoire de Physique Th\' eorique, Ecole Normale Sup\' erieure, 
24 Rue Lhomond, 75231 Paris Cedex 05, France.}
\vskip 0.5cm

\abstract {
Higher dimensional generalisations of self-duality conditions and of theta angle terms are analysed
in Yang-Mills theories. For the theory on a torus, the torus metric and various 
antisymmetric tensors are viewed as coupling constants related by U-duality, arising from background 
expectation values of
supergravity fields for D-brane  or matrix theories.  At certain special points in the moduli space
of coupling constants certain branes or instantons are found to dominate the functional integral.
The possibility of lifting chiral or supersymmetric theories to higher dimensions is discussed.
}

\endpage

\chapter{Introduction}

Supersymmetric Yang-Mills (SYM) theories in $D$ dimensions play a crucial role in the study of 
D-branes and in the matrix approach to M-theory.
The dynamics of a Dirichlet $p$-brane is described by a low-energy effective action 
 for the  SYM multiplet in $D=d+1$ dimensions (obtained by reducing from $D=10$)
which is a Born-Infeld action plus couplings to RR gauge fields through a Wess-Zumino term
[\Dbrane,\polch]. The matrix theory conjecture  [\Mat,\brev] relates M-theory compactified on a $d$
torus
$T^d$ to SYM in
$D=p+1$ dimensions on $\R\times \ti T^d$ where $\ti T^d$ is the dual torus [\Mat-\mem]. For $d\ge
4$ the SYM is not renormalizable and extra degrees of freedom are needed at high energies, but the
SYM is still a useful effective description for many purposes. 

In $D=4$, the addition of a topological $\th$-angle term $\th F^2$ to the $N=4$ SYM lagrangian
led to an enlargement of the Montonen-Olive duality to $SL(2,\Z)$, which was the key to many later
developments in the study of duality. The angle $\th$ is a coupling constant of the SYM which arises
from  string theory as the expectation value of a certain field.
 The D-brane action for a $p$ brane is a $D=p+1$ dimensional action including the following terms governing the world-volume YM fields
$$S=\tr   \int  \left[ {1\over g^2}   {\cal F}  \lll *{\cal F} + C_D +C_{D-2}  {\cal F} 
+C_{D-4} {\cal F}^2
+C_{D-6} {\cal F}^3
+... 
+C_{D-2r}  {\cal F}^r \right ]
\eqn\Dact$$
where 
$$ {\cal F} _{mn} =F_{mn} -B_{mn} \unit ,
\eqn\abc$$ 
$F_{mn}$ is the YM field strength, $r$ is the integer part of $D/2$,
$B_{mn}$ is the NS-NS 2-form gauge field
 and the $C_m $ are $m$-forms arising from the background
expectation values of RR gauge fields [\Dbrane].
Similar actions    arise in matrix theory. 
From the  point of view of the SYM theory, the  forms $C_m$ are again coupling constants.
As will be discussed elsewhere [\prog],
 including  terms such as these is necessary if there is to be an
   enlargement of the  expected $SL(d,\Z)$ symmetry of SYM on $\R\times \ti T^d$ to the appropriate
U-duality group for
$d>3$, as has been found to be the case for $d=3,4,5$. The moduli space  for SYM on $\R\times \ti
T^d$ includes the moduli space
$\R\times SL(d)/SO(d)$
 of metrics on $\ti T^d$, together with the coupling constants arising from constant values of the
forms $C_m$,  and the U-duality group acts on this space, mixing the torus metric with the various
anti-symmetric tensor gauge fields [\prog]. This generalises the way that including the
$\th$-angle for
$d=3$ leads to the U-duality group
$SL(3,\Z)\times SL(2,\Z)$; in this case the $\th$-angle is the $SL(2,\Z)$ partner of the coupling
constant $g$, or torus volume. More generally, the forms $C_m$ are the U-duality partners of the
torus metric, so that it is necessary to include such couplings to understand U-duality [8].

In SYM and D-brane actions (in planar gauge), there are  adjoint-valued scalar fields $X^i$
($i=1,\dots, 10-D$) taking values in a transverse space. Then the general \lq topological' term 
in $D$ dimensions
can
involve $dX$, giving terms
$$\sum _n  Tr(Y_{D-2n}  F^n)
\eqn\abc$$  where 
$$Y_m= \sum_p K_{i_1...i_p} DX^{i_1}... DX^{i_p}Z_{m-p}, 
\eqn\abc$$ where $Z_m$ is an  $m$-form on the D-dimensional space and
$$
DX^i = dX^i +[A,X^i]
\ek
 Thus the   action is
parameterised by space-time forms $Z_m$ and by the transverse forms $K$. 
The dimensional reduction of terms proportional to $\tr F^n$ gives terms
involving $\tr [ (DX)^{n-m} F^m]$ in $D=n+m$ dimensions.
Such terms can play an important role in SYM and will be discussed further in [\prog].

Our purpose here is to study some of the consequences of including such topological terms in the
SYM action, and in particular the instantons or solitons that dominate the functional integral.
While there has been considerable interest in such terms for special choices of the
forms $C_m$, such as the covariantly constant forms on manifolds of special holonomy
[\Corr-\Bobb], our viewpoint here is rather different, as we wish to consider the theory as a
function of these coupling-tensors, and consider the properties of SYM as these vary. 
The set of coupling constants or moduli of the SYM  on some space-time $M$   then include the 
moduli of metrics on $M$ and the forms $C_m$ on $M$ (which arise from string background fields,
and will usually be taken to satisfy the classical field equations). In particular,  there can be solitonic $p$-brane
solutions of the SYM which couple to the
$p+1$ form
$C_{p+1}$ and which are interpreted as $p$-branes in the matrix theory; for example, in 5+1 dimensions,
solitons coupling to the  2-form $C_2$ correspond to strings, and the matrix model is in fact a
(non-critical) string theory [\brs,\dvv,\MatTor], while in 6+1 dimensions the matrix theory has
membrane excitations [\mem]. The
$d+1$ dimensional SYM corresponding to M-theory on $T^d$ then contains $d-4$ branes for $d \ge 4$.

We shall particularly interested in  
the quadratic YM Lagrangian in $D=d+1$ dimensions involving a 4-th rank tensor
$X_{mnpq}$, 
$$ {1\over 4g^2}\tr    F^{mn}F_{mn} +{1\over 4}X^{mnpq}\tr   F_{mn} F_{pq}
\eqn\ymt$$ 
which 
arises from the quadratic terms in \Dact, with $X\propto *C_{D-4}$.
This
depends on the following coupling constants or moduli: the D-dimensional metric,
the YM coupling $g$ (which can be absorbed into the metric) and a 4-form
$X^{mnpq}$. This is always  part of the low-energy limit of the matrix theory for M-theory on
$T^{d}$. On a curved space, the second term is topological (if, as we shall assume, $d*X=0$, so
that the action depends only on the cohomology class of $*X$)  and gives a generalised
$\theta$-angle; a different $\theta$-angle arises for each homology  4-cycle [\Dis]. Such terms were
considered in the context of matrix models in [\Dis].

We can generalise this action to allow an
   $X$ that is not   a totally antisymmetric tensor, but is a more general 4-th rank tensor
satisfying
$$X_{mnpq}=
-X_{nmpq}=
-X_{mnqp}=
X_{pqmn}
\eqn\genx$$
For example, in the D-brane action, including the NS-NS 2-form $B$ gives an action
\ymt\ with
$X_{mnpq}=B_{mn}B_{pq} + *C _{mnpq}+...$ where $*C$ is the dual of the RR $D-4$ form potential
$C_{D-4}$.

It will be important in what follows that the \lq topological term' in \ymt\
 can sometimes be real
in the Euclidean action, unlike the usual $D=4$ theta-angle term, which is imaginary.
The action appearing in the functional integral is the Euclidean one resulting from the 
Wick rotation $t \to it$.
In $D=4$, the Minkowski space term $\theta \int\tr   F\lll F$ with real $\theta$ becomes
$i\theta \int\tr    F\lll F$ in Euclidean space, so that $\th$ is an angle, coupling to the second
Chern class. In any dimension, the Wick rotation $t \to it$ is accompanied by $A_t \to -i A_t$
so that the electric fields $E_i \equiv F_{0i} $ are rotated $E_i\to -i E_{i}$ while the magnetic
fields $B_{ij}\equiv F_{ij} $ are unchanged, $B_{ij} \to B_{ij}$.
In Minkowski signature, 
  the lagrangian \ymt\
should be real so that the coupling constants $X^{mnpq}$ are real.
On Wick rotating, the action \ymt\
 becomes
$$ {1\over 2g^2}\tr   \left( E_iE^i+ \2 B^{ij}B_{ij} \right)
+{1\over 4}X^{ijkl}\tr   B_{ij} B_{kl} + i {1\over 2}X^{0ikl}\tr   E_{i} B_{kl}
\eqn\ymte$$ 
Thus the coefficient of $E\lll B$ becomes imaginary (as for the usual 4-dimensional $\th$ angle)
while that of $B\lll B$ remains real.
Thus the $X^{0ikl}$ become angular variables (for fixed $i,j,k$) while the
$X^{ijkl}$ will not satisfy any periodicity conditions in general.
We shall be interested in 
embedding an $n$-dimensional instanton into a  $d+1$ dimensional Lorentzian space
$(n \le d)$ and the 
couplings
 $X_{mnpq}$ with purely spatial indices that contribute to the instanton action real on Wick
rotating.


\chapter{Instantons Satisfying a Generalised Self-duality Condition}

Consider configurations satisfying a generalised self-duality equation
$${ 1\over 2}Y_{mnpq}F^{pq} =\lambda F_{mn}
\eqn\sdu$$
  for some  4-form $Y$ and constant $\ll$. These  will play an important  role when $X \propto Y$.
The Bianchi identity   implies that a configuration satisfying \sdu\ also satisfies the field
equation
$D^mF_{mn}=0$.
(Note that this would no longer be true if $Y$ were not totally 
anti-symmetric, and was replaced by a
tensor with the symmetries \genx.)
In this section we will consider instanton solutions to \sdu\ in Euclidean space, and will embed
these in higher dimensional Minkowski spaces to obtain brane solutions in the next section.
An alternative generalisation of the self-dual YM to $D>4$ dimensions was proposed in [\Der].

 Instanton solutions to \sdu\ have been studied in the case in which $Y$ is invariant under
a subgroup $SU(n)$, $G_2$ or $Spin(7)$ of the Lorentz group in flat space, and
in the case of manifolds of holonomy $SU(n)$, $G_2$ or $Spin(7)$
with the tensor $Y$ covariantly constant. We shall  consider here the case of flat space-time
and constant tensors $Y,X$.
In 4 Euclidean dimensions, $Y$ is proportional to the volume-form
 and solutions satisfying \sdu\
(with $\ll$ given by $1$ or $-1$ if $Y$ is conventionally normalised) are  
 self-dual or anti-self-dual instantons on
$N$ satisfying \sdu. In 8 Euclidean dimensions, if 
$Y$ is the $Spin(7)$ invariant
self-dual 4-form, then there are point-like instantons
satisfying \sdu\    [\SpinSdSol,\Pop] (with $\ll$ chosen so that  $F$ is projected
into the {\bf 21} of
$Spin(7)$).
 Similarly, in 7 dimensions, if  $Y$ is invariant under $G_2$,  there are  pointlike instantons
satisfying \sdu\   [\Pop,\Spingt]. Finally, in $2m$ dimensions, if  $Y$ is invariant under $SU(m)$,
then the action is extremised by instantons satisfying \sdu. The Yang-Mills field   is then a
connection of a holomorphic vector bundle satisfying the Uhlenbeck-Yau equation, and  point-like
instantons are again expected. Similar instantons in 6 dimensions were considered in [\Tset].

 For the $Spin(7)$ and $G_2$
solutions in $\R ^8$ or $\R^7$, the 
  YM action
$$\int \vert F\vert^2
\eqn\abc$$
is infinite  because of the slow fall-off of the fields.
If however, there are similar instanton solutions    on a compact space, such as a torus, then it is
conceivable that the action could be finite in that case.

  The tensors $X^{mn,pq},Y^{mn,pq}$ can both be
regarded as  
   $N\times N$ 
symmetric matrices ($N=D(D-1)/2$)
 whose rows and columns are labelled by index pairs $mn$, $pq$ respectively.
It will be convenient to denote these matrices as $X_{ab},Y_{ab}$ respectively, where $a,b=1,...,N$.
General tensors $X_{mnpq}$ satisfying \genx\ will correspond to matrices $X_{ab}$ with $N$
independent eigenvalues, while requiring 
$X_{mnpq}$ to be totally anti-symmetric imposes 
constraints on these eigenvalues, and in particular that $X_{ab}$ is traceless.
The $SO(4),Spin(7)$ and $G_2$ cases considered above are ones in which  $Y$
satisfes a quadratic characteristic equation. 
In general, $Y$ will have $N$ real eigenvalues $\ll_a $ (not necessarily distinct), 
so that the 
kinetic term
 can be written as
$$
\int \tr   \sum _{a=1}^{N}  {1\over g^2}F^aF^a
\eqn\actdi$$
after writing $F_{mn}$ as a $D(D-1)/2$ dimensional vector 
and transforming to the (orthonormal frame) basis in which 
the kinetic term becomes $g^{-2} \sum _a F^aF^a $
(after a rescaling of $g$) and  in which
$Y_{ab}$ is diagonal, $Y_{ab}= diag(\ll_1,\dots \ll_N)$.
Note that the total anti-symmetry of $Y_{mnpq}$ implies that $Y_{ab}$ is traceless.

It follows   that the kinetic term $\int F\lll*F$ is bounded below by a term
proportional to the topological term, since, in the basis in which  $Y_{ab}= diag(\ll_1,\dots \ll_N)$,
$$ \tr   \sum _a (F^a)^2 =tr \sum _{a,b} {1\over \ll_{a}} Y_{ab}F^aF^b
 \ge {1\over \ll_{max}}tr \sum _{a,b} Y_{ab}F^aF^b
\eqn\abc$$ 
implies
$$\int
tr \vert F \vert ^2\ge   {1\over \ll_{max}} 
\int (*Y) \lll tr( F \lll F)
\eqn\abc$$
where $\ll_{max}$ is the largest of the eigenvalues $\ll_a$.
As $\sum _a \ll_a =0$, the minimum eigenvalue is negative, $\ll_{min}=-\mm$, $\mm>0$, and
a similar argument implies
$$\int tr
(F\lll*F) \ge  -{1\over \mm} 
\int (*Y) \lll tr(F \lll F)
\eqn\abc$$
The first   bound is saturated if $F$ satisfies the self-duality condition \sdu\ with eigenvalue
$\ll_{max}$, while the second is saturated
if $F$ satisfies the self-duality condition \sdu\ with eigenvalue
$\ll_{min}$.
For any self-dual $F$ satisfying \sdu\ for some $\ll$, the kinetic term is
proportional to  the topological term
$$  tr( F\lll*F) =    {1\over \ll{}}  (*Y) \lll\tr   (F \lll F)
\eqn\abc$$ 

Consider now the action \ymt\ with $X= \th Y$, 
$$S={1\over g^2}\tr   \int d^Dx \, F^2 +{1\over 2}\int d^Dx \, \th Y^{mnpq}\tr (  F_{mn} F_{pq})
\eqn\ymth$$ 
which becomes
$$S=\int   \sum _a  \left({1\over g^2}+\th \ll_a\right)\tr   (F^a)^2
\eqn\abc$$
The action will be positive-definite if the eigenvalues of $\dd _{ab}+g^2 X_{ab}$ are all greater
than zero, and this will clearly be the case for small enough coupling $g$.
At large coupling, the SYM description will break down for $d>3$ 
(for $d=3$, the strong coupling limit is described by a dual SYM theory, for $d=4$ an extra
dimension emerges to give   a 5+1 dimensional self-dual tensor theory etc)  and the semi-classical
analysis is in any case not applicable.

Taking $F$ to satisfy
  the self-duality condition \sdu\ with
$\ll$ given by
$\ll =- \ll_a$ for any of
  these eigenvalues will give a stationary point of the action.
Choosing $\th=-g^{-2}/\ll_{max}$ gives the action
$$S=\int \tr   \sum _a  {1\over g^2 \ll_{max}}
\left(\ll_{max}- \ll_a\right)(F^a)^2
\eqn\abc$$
which is positive and vanishes for self-dual solutions satisfying \sdu\ with $\ll=\ll_{max}$.
The semi-classical functional integral is dominated at weak
coupling by those solitons with zero action; all others are suppressed by factors of
$\exp {(-1/g^2)}$.
Thus with this choice of action, with the topological term given in this way in terms of $Y$, the
weakly-coupled theory is dominated by the instantons that are $Y$-self-dual \sdu\ with eigenvalue
$\ll_{max}$. 
Similarly, choosing
 $\th=g^{-2}/\mm$ gives the action
$$S=\int \tr   \sum _a  {1\over g^2 \mm}
\left( \ll_a - \ll_{min}\right)(F^a)^2
\eqn\abc$$
which is positive and vanishes for self-dual solutions satisfying \sdu\ with $\ll=\ll_{min}$, and
these would dominate at weak coupling.

\chapter{$p$-Brane Solutions and Supersymmetry}

Consider solutions of \sdu\ in a $D=d+1$ dimensional flat space with Lorentzian signature. If $N$ is
an $n$-dimensional Euclidean submanifold and there is an instanton solution on $N$ satisfying \sdu\
for some 4-form $Y$, then this will lift to a $p$-brane solution in $D$ dimensions with
$p=d-n$.\foot{The Yang-Mills connection
is independent of the coordinates transverse to $N$ and the components transverse to $N$ vanish.}
 For example, a 4-dimensional instanton leads to a  0-brane in
5-dimensions or a
 string in 6-dimensions.
 The theory with action \ymt\ has saddle points corresponding to all self-dual solutions that satisfy
\sdu\ for some $Y$ and some $\ll$ (which must be an eigenvalue of $Y$ for a non-trivial solution).
Thus the theory will have BPS $p$-brane solutions with $p=d-4$ (if $d \ge 4$),
with $p=d-7$ (if $d \ge 7$),
and with $p=d-8$ (if $d \ge 8$),
 corresponding
to  4-dimensional
$SU(2)$ instantons,  7-dimensional $G_2$ instantons and 8-dimensional $Spin(7)$ instantons,
respectively. 

It was seen in the last section that, by choosing $X$ to
 be proportional to   $\th Y$ with   appropriate tuning of the coefficient $\th$,
one can arrange
for precisely one type of self-dual instanton (those self-dual with respect to $Y$ with either
maximum or minimum eigenvalue) to have zero action and hence to dominate the path integral. 
This 
can be lifted to the
 $p=d-n$ brane solutions;  if the pull-backs of $X$ and
$Y$ to $N$ agree, $X \vert_N = \th Y \vert_N $, with   appropriate choice of $\th$, then
the $p$-branes satisfying \sdu\ with either
maximum or minimum eigenvalue will have zero transverse action (i.e. action per unit $p$-volume).
Here it is important that the term in the action
\ymt\ involving $X
\vert_N$ remains real; note that both the Euclidean and Lorentzian actions vanish in this case.
Thus there are points in the SYM moduli space (corresponding to special choices of
$X$) at which certain types of brane have zero action, even though they will in general have 
non-zero energy densities. Note that this is true for any value of the coupling
 $g$, and is a different phenomenon from the
behaviour at strong coupling.
Thus for special choices of $X$, a class of branes of a certain orientation is \lq selected' to have
zero action and so to dominate the functional integral, especially at weak coupling, when other
branes are suppressed. One possible interpretation of this might be that at such points the vacuum
is modified by a condensation of a particular class of
$p$-brane.

For example, self-dual instantons on a 4-dimensional submanifold $N$ give rise to BPS $d-4$ branes
in $d+1$ dimensions with finite energy density, proportional to $1/g^2$. At   points in moduli
space at which the pull-back of $X$ to $N$  is $-g^2$ times the volume form on $N$, the
 action of these $d-4$ branes vanishes, while the transverse action of all other $p$-branes
(such as $d-4$ branes associated with other 4-submanifolds) remains of order $1/ g^2$ in general.
For $d=4$, these 0-branes become light at strong coupling, and the strong-coupling limit
corresponds to a decompactification to 5+1 dimensions [\rozali] with the 0-branes interpreted  as
Kaluza-Klein modes. 
For $d=5$, these solitonic branes are the strings of the non-critical string theory, for $d=6$ these
are membranes etc. Their presence is reflected by the presence of 
a $d-4$ form \lq central' charge in the $d+1$ dimensional superalgebra.
However, the instantons on $N$  only have zero action if $X$ is the volume form of $N$.

 The
$d-4$ branes are BPS and have finite action and energy density for all values of $Y$, but their
action becomes zero  for the special choice of $X \propto Y$. The $d-7,d-8$ branes are (formally) 
BPS, but their total energy   is   infinite, as is the action for all values of $X$ except the
special value at which the action vanishes.

The $Spin(7)$ and $G_2$ instantons in $\R^8$ or $\R^7$  respectively have infinite
action, and so
 these and the corresponding
$d-7,d-8$ branes will be infinitely  suppressed in the functional integral.
 Choosing $X$ to be proportional to some $Y$ with the appropriate coefficient will arrange
for precisely one type of self-dual solution 
to have an action that is formally zero (the integrated kinetic and topological terms are separately
divergent, but the Lagrnagian densities cancel). The energy per unit $p$-volume will
remain divergent, so that the interpretation in this case is unclear.
However, the   actions for the instantons
in $\R^8$ or $\R^7$, and the 
corresponding brane actions and energies,
 are infinite because of the slow fall off of the solution, and it
would be interesting to see whether there are similar solutions on a  torus (or compact space of
special holonomy) and whether such solutions have finite action.
If there were such finite action instantons on
$T^8$ or $T^7$, they could play an important role in the matrix models for M-theory compactified on
$T^d$ for $d\ge 7$. For $d=7,8$, there would be a 0-brane in $d+1$ dimensions that became light
at strong coupling,  which could be related to a decompactification to one higher dimension, as
in the case of $d=4$. This possibility will be discussed further elsewhere [\prog].

Thus, at least for weak coupling, the functional integral has saddle-point solitons satisfying the 
generalised self-duality equations \sdu, and
  it is clearly important to understand the properties of solutions to \sdu, and in
particular whether they are point-like or brane-like.  The spectrum of solutions would then
determine the brane-spectrum of \ymt.
As the topological term is topological, the classical solutions of the theory are the same for all
values
of the coupling $X_{mnpq}$, but changing $X$ changes the 
weight corresponding to each in the semi-classical approximation, and changes the
subset of solutions that dominate the
functional integral.
In particular, the presence of $X$  breaks the Lorentz group down to
the sub-group preserving $X$, and for special choices  of  $X$ (corresponding to special
points in the moduli space) the Lorentz symmetry is \lq enhanced' and there is the possibility of the
spectrum  of branes for which the action vanishes
also
being enhanced.
In such cases, it is often possible to twist the SYM to obtain a topological field theory [\Bob].


We consider now the supersymmetry of configurations satisfying \sdu. 
In SYM, the   supersymmetry transformation of the spinor field $\chi$ is
$$\dd \chi = \2 F_{mn}\ggg^{mn} \ee +...
\eqn\varr$$
where the ellipses refer to   extra terms involving  scalar fields.
For configurations involving only the YM fields (i.e. with vanishing scalar fields) and which
satisfy  
 \sdu, the variation \varr\ will vanish for spinorial parameters $\ee$ satisfying
$$
\ll \ggg_{mn}\ee = -\2 Y_{mnpq}
\ggg^{pq}\ee
\eqn\abc$$
This implies that $\ee$ satisfies
$$
(\unit - \aa _\ll \ggg )\ee=0
\eqn\chir$$
where
$$
\ggg = {1\over 4!} Y_{mnpq}\ggg ^{mnpq}
\eqn\giss$$
and 
$$ \aa _\ll=
{12 D(D-1) \over \ll}
\eqn\abc$$
Thus a solution to \sdu\ with a particular value of $\ll$ will be preserved under those
supersymmetries whose parameters satisfy the chirality constraint \chir.

\chapter{Antisymmetric Tensor Gauge Theories}

This can be generalised to other fields.
For a 2-form gauge theory with $H=dB$, the action \ymt\ generalises to
$$ {1\over g^2} \vert H^2 \vert + {1\over 72}D^{mnpqrs} H_{mnp} H_{qrs}
\eqn\abc$$
where $D^{mnpqrs}$
 is a tensor with the symmetry properties
$$D^{mnpqrs}=D^{[mnp]qrs}=D^{mnp[qrs]}=D^{qrsmnp}
\eqn\abc$$
Note that although it cannot be totally anti-symmetric in this case,
it could be taken to be the \lq square' of a 6-form $ X^{mnpqrs} $, with
$$ D^{mnpqrs} = \6
X^{mnptuv} X_{tuv}{}^{qrs} 
\eqn\dfac$$

Consider  the generalised self-duality equations 
$$ H_{mnp}= {\aa\over 6} Y_{mnpqrs}H^{qrs}
\eqn\hdu$$
for some tensor $Y_{mnpqrs}$, which will be assumed to be totally antisymmetric so that the Bianchi identity implies the
field equation for $H$.
The 6-form  $Y_{mnpqrs}$ can be regarded as an anti-symmetric matrix in the triplets of indices
$mnp$ and $qrs$ and can be skew-diagonalised with eigen-values $\pm \aa_a$, or alternatively
diagonalised over the complex variables with complex conjugate eigenvalues.
Defining
$$ C^{mnpqrs} = \6
Y^{mnptuv} Y_{tuv}{}^{qrs} 
\eqn\dfaca$$
 the tensor  $C_{mnpqrs}$ can be regarded as a symmetric matrix in the triplets of indices
$mnp$ and $qrs$ and can be diagonalised, with eigenvalues $\ll _a = \aa _a ^2$. 
The self-duality condition \hdu\ implies
$$ H_{mnp}= {\ll\over 6} C_{mnpqrs}H^{qrs}
\eqn\cdu$$
  with
$$\ll = \aa^2
\eqn\abc$$
As in the 2-form case, the $H_{mnp}$ can be decomposed into eigenstates of $C $, and there is a bound on the kinetic term
corresponding to the largest and smallest eigenvalues $\ll_{max}, \ll_{min}=-\mm$:
$$ \int \vert H \vert ^2 \ge {1 \over \ll_{max}} \int (*Y )\lll H \lll *[(*Y )\lll H]
\eqn\abc$$
and
$$ \int \vert H \vert ^2 \ge -{1 \over \mm} \int (*Y )\lll H \lll *[(*Y )\lll H]
\eqn\abc$$
These bounds will be saturated if $H$ satisfies the self-duality conditions \hdu\ with $\aa = \pm
\sqrt \ll_{max}$ or $\aa = \pm \sqrt  \ll_{min}$.
Again, by considering the action with \lq topological' term \ymt, we can arrange for the action to
vanish for these self-dual solutions by choosing  $D\propto C$ with an appropriate constant of
proportionality.

In a supersymmetric tensor multiplet   there is a spinor  transforming as
$$\dd \chi = \6
H_{mnp} \ggg ^{mn p} \ee +...
\eqn\abc$$
where the ellipses refer to   extra terms involving   scalar fields and fermion bilinears.
For configurations involving only the YM fields and which satisfy  
 \hdu, \varr\ will vanish for spinorial parameters $\ee$ satisfying
$$
(\unit - \beta _\aa \ggg )\ee=0
\eqn\chirh$$
where
$$
\ggg = {1\over 6!} Y_{mnpqrs}\ggg ^{mnpqrs}
\eqn\yisss$$
for some $\beta_\aa$.
Thus a solution to \hdu\ with a particular value of $\aa$ will be preserved under those
supersymmetries whose parameters satisfy the chirality constraint \chirh. 

\chapter{Chirality and Self-Duality in Higher Dimensions}

In this paper we have considered theories whose
\lq coupling constants' include background tensors; in the case of 
Yang-Mills theories, the coupling constants
included $g$, the metric $g_{mn}$ and a 4-form $X_{mnpq}$. In D-brane actions and matrix theories,
these emerge from the expectation values of certain fields. In particular,  the matrix theory
for M-theory on $T^d$ is related to SYM on $\ti T ^d\times \R$ and in  this context it is natural
to consider the metric on $\ti T ^d$ and the expectation values of various tensor gauge fields on
 $\ti T ^d$ as coupling constants of the matrix theory.
Given such non-Lorentz-invariant coupling constants, it is possible to generalise the notions of
chirality and self-duality to higher dimensions, albeit in a rather trivial way, and also to obtain
supersymmetric theories in higher dimensions. 
(This is related to the work of [\sez], in which theories with extra constant vectors can be
supersymmetric in more than 11 dimensions; in the present context, such vectors could be thought of
as coupling constants.)

For example, given a 4-form coupling constant $Y_{mnpq}$ in $D$ dimensions, one can define
generalised
  self-dual YM fields through \sdu\ and generalised chiral spinors by
$$ \ggg \ll = \ll
\eqn\fchi$$
where $\ggg$ is given by \giss.
In 4 Euclidean dimensions, there is a supersymmetric system consisting of  self-dual YM
coupled to a chiral fermion [\SSDU], and
it is possible to generalise this system to a higher dimensional supersymmetric system in this 
way.
For example, in 4+1 dimensions (with signature $(+,-,-,-,-)$), the 4-form $Y$ is dual to a vector 
$V$
and
\sdu\ becomes
$$*F = V \lll F
\eqn\abc$$
which implies that $V^mF_{mn}=0$ and $V^2 F=F$, so that $F=0$ unless the vector  $V$ is
time-like with $V^2=1$, in which case $A_m$ is independent of time in the gauge $A_0=0$ and the
YM sector reduces to 4-dimensional Euclidean self-dual YM.
The fermion chirality constraint \fchi\ then implies
$V_m \ggg ^m \ll= \ll$ which, together with the Dirac equation, implies that the spinor reduces to
a chiral spinor in  4-dimensional Euclidean space. Thus the theory reduces to the supersymmetric 
self-dual YM system in 4 Euclidean dimensions. In higher dimensions, similar results should apply
whenever the 4-form $Y$ is the volume-form for a 4-dimensional Euclidean submanifold.

In a similar way, it is possible to lift the 6-dimensional self-dual tensor theory to higher
dimensions.  In 5+1 dimensions, there is a (2,0) supersymmetric theory of a 2-form whose field
strength $H$ satisfies a self-duality constraint, together with a chiral spinor and 5 scalars.
This could be lifted to $D>6$ dimensions using a 6-form $Y$ to a define generalised self-duality
constraint on $H$ \hdu\ and a generalised chirality constraint on the spinors, projecting onto
aparticular eigenvalue of the chirality operator \yisss. If
$Y$ is
$SO(5,1)$ invariant, so that it corresponds to the volume form on a $5+1$ dimensional submanifold,
then the lifted theory should again   be supersymmetric.

\ack
{I would like to thank Bobby Acharya, Mike Douglas, Jos\' e Figueroa-O'Farrill, Jerome Gauntlett,
Bernard Julia and Steve Thomas for useful discussions.}

\refout

\bye